\documentclass[]{spie}  

\usepackage{amsmath,amsfonts,amssymb}
\usepackage{graphicx}
\usepackage[colorlinks=true, allcolors=blue]{hyperref}
\usepackage{subcaption}
\usepackage{multirow}
\usepackage{caption}

\usepackage[ruled,vlined]{algorithm2e}
\SetAlFnt{\small}  
\RestyleAlgo{boxruled}

\usepackage[dvipsnames]{xcolor}
\usepackage{tcolorbox}
\definecolor{light_yellow}{RGB}{254, 224, 145}
\definecolor{light_orange}{RGB}{254, 220, 200}
\definecolor{light_blue}{RGB}{160, 201, 255}
\definecolor{light_green}{RGB}{213, 231, 211}
\definecolor{light_purple}{RGB}{231, 223, 246}
\definecolor{mrtrix_blue}{RGB}{0, 122, 255}
\definecolor{torch_red}{RGB}{225, 0, 0}

\title{Fully Differentiable dMRI Streamline Propagation in PyTorch}

\author[a]{Jongyeon Yoon}
\author[b]{Elyssa M. McMaster}
\author[a]{Michael E. Kim} 
\author[b]{Gaurav Rudravaram}
\author[c,d]{Kurt G. Schilling}
\author[a,b,c,d]{Bennett A. Landman}
\author[a]{Daniel Moyer}
\affil[a]{Department of Computer Science, Vanderbilt University, Nashville, TN, USA}
\affil[b]{Department of Electrical and Computer Engineering, Vanderbilt University, Nashville, TN, USA}
\affil[c]{Department of Radiology and Radiological Sciences, Vanderbilt University, Nashville, TN, USA}
\affil[d]{Vanderbilt University Institute of Imaging Science, Vanderbilt University, Nashville, TN, USA}

\begin{document} 
\maketitle

\begin{abstract}
Diffusion MRI (dMRI) provides a distinctive means to probe the microstructural architecture of living tissue, facilitating applications such as brain connectivity analysis, modeling across multiple conditions, and the estimation of macrostructural features.
Tractography, which emerged in the final years of the 20th century and accelerated in the early 21st century, is a technique for visualizing white matter pathways in the brain using dMRI.
Most diffusion tractography methods rely on procedural streamline propagators or global energy minimization methods.
Although recent advancements in deep learning have enabled tasks that were previously challenging, existing tractography approaches are often non-differentiable, limiting their integration in end-to-end learning frameworks.
While progress has been made in representing streamlines in differentiable frameworks, no existing method offers fully differentiable propagation.
In this work, we propose a fully differentiable solution that retains numerical fidelity with a leading streamline algorithm.
The key is that our PyTorch-engineered streamline propagator has no components that block gradient flow, making it fully differentiable.
We show that our method matches standard propagators while remaining differentiable.
By translating streamline propagation into a differentiable PyTorch framework, we enable deeper integration of tractography into deep learning workflows, laying the foundation for a new category of macrostructural reasoning that is not only computationally robust but also scientifically rigorous.
\end{abstract}


\section{INTRODUCTION}
\label{sec:intro}  
Diffusion MRI (dMRI) enables unique characterization of microstructural enviroments within living tissue, supporting analyses such as brain connectivity mapping, multi-condition modeling, and inference of macrostructural properties.
Among its diverse applications, dMRI streamline tractography is widely recognized as the premier method for non-invasively mapping white matter (WM) tracts/fasciculi in the brain, where generated streamlines are used in both surgical planning and clinical research for neurological disorders \cite{jones2010diffusion, essayed2017white, kamagata2024advancements}.
The most common approach for obtaining tracts typically begins with estimating fiber orientation distributions (FODs) from raw dMRI signals, often via constrained spherical deconvolution (CSD) \cite{tournier2004direct, tournier2007robust}. 
These FODs are then used in conjunction with streamline propagation algorithms, such as SD\_Stream \cite{tournier2012mrtrix}, to trace out WM tracts.

Recent advances in deep learning enabled tasks that were previously difficult or infeasible, such as predicting streamlines from structural T1-weighted MRI \cite{cai2023convolutional, yoon2025tractography, yoon2025transformer}. 
However, a major limitation has been that conventional streamline propagators are inherently non-differentiable, limiting their direct integration into end-to-end learning frameworks.
While there has been progress in differentiable representations of streamline data, no existing method has achieved fully differentiable streamline propagation.

In this work, we introduce a fully differentiable streamline propagator implemented in PyTorch that maintains numerical fidelity with the leading streamline algorithm, SD\_Stream from the MRtrix package \cite{tournier2012mrtrix}, while remaining compatible with gradient optimization.
Rather than designing a new model, we translated the SD\_Stream algorithm in MRtrix, which was originally implemented in C++, into PyTorch.
The key innovation is that our PyTorch-engineered streamline propagator contains no components that block the gradient flow, making it fully differentiable. 
We show that our differentiable PyTorch streamline propagator reconstructs streamlines that closely match those generated by SD\_Stream, achieving agreement within floating-point precision, as quantified by the Hausdorrff distance between tracts produced by both methods from identical data.
Furthermore, we show that our streamline propagator enables gradient computation, with reasonable computation time and memory consumption.

Our results demonstrate that our differentiable streamline propagator enables integration of streamline data into end-to-end deep learning pipelines, establishing a foundation for a new class of macrostructural reasoning that is both computationally robust and scientifically rigorous.

\begin{figure}[ht]
    \centering
    \begin{subfigure}{0.4\textwidth}
        \centering
        \includegraphics[width=\linewidth]{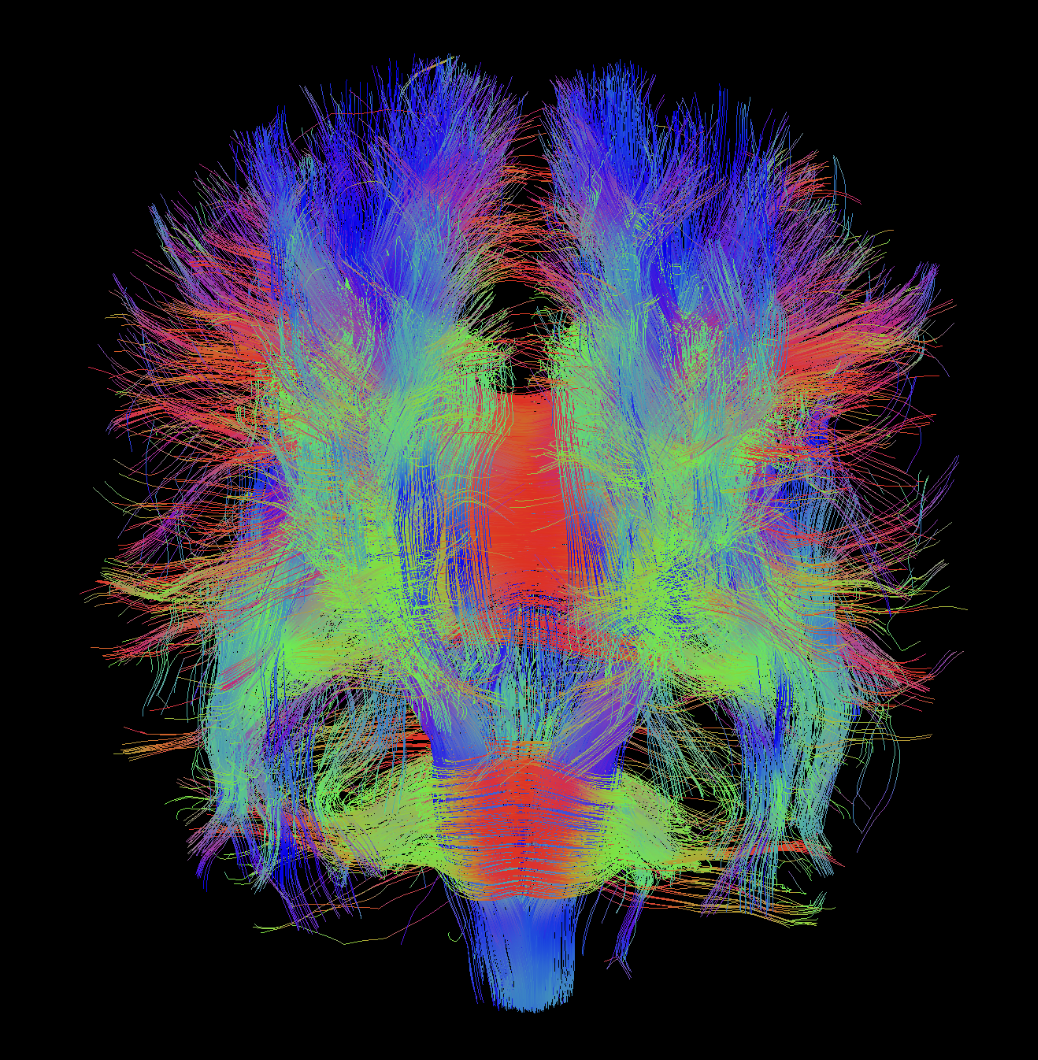}
        \label{fig:subfig1}
        \subcaption{Streamlines from \textbf{MRtrix}}
    \end{subfigure}
    \begin{subfigure}{0.4\textwidth}
        \centering
        \includegraphics[width=\linewidth]{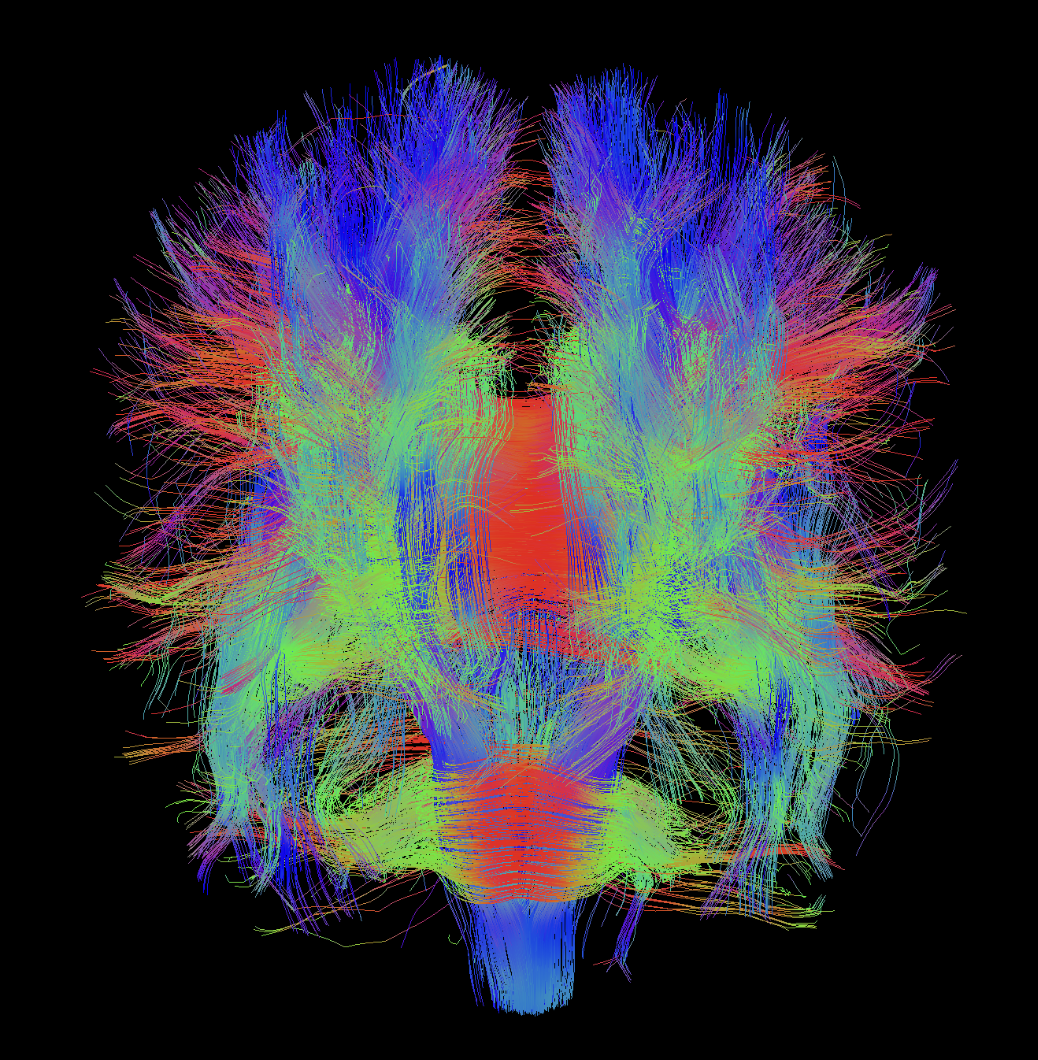}
        \label{fig:subfig2}
        \subcaption{Streamlines from \textbf{PyTorch}}
    \end{subfigure}
    \caption{MI-Brain\cite{rheault2016mi} visualization of 10,000 streamlines from same FODs using SD\_Stream from MRtrix (left) and PyTorch (right). Given the same seed points and initial propagation directions, differentiable PyTorch streamline propagator reproduces streamlines that are visually indistinguishable from MRtrix.}
    \label{fig:whole_brain}
\end{figure}

\section{Methods}

\subsection{SD\_Stream}
\label{sec:mrt_SDS}
SD\_Stream \cite{tournier2012mrtrix} is a deterministic streamline tractography algorithm provided by the MRtrix software package, implemented in C++.
It propagates streamlines by iteratively stepping along the peak fiber orientation most aligned with the current tracking direction at each point.
Tracking is initialized from seed points $S = \mathbf{x}_0$, where $\mathbf{x}_0$ is a position randomly generated by uniform sampling within a user-specified seed region-of-interest (ROI) mask, and randomly assigned unit vectors $\mathbf{d}_0$, sampled uniformly on the sphere to define initial propagation direction.

Since diffusion MRI data stores FODs at discrete voxel locations, while seed points can exist at arbitrary locations in continuous 3D space, the algorithm first estimates the FOD coefficients at each seed location by performing trilinear interpolation:
\[
c(\mathbf{x}) = \sum_{i,j,k} w_{ijk} c_{ijk},
\]
where $w_{ijk}$ are the trilinear interpolation weights for voxel $(i,j,k)$ surrounding $\mathbf{x}$.
The interpolated FOD at position $\mathbf{x}$ is represented in the spherical harmonic (SH) basis as:
\[
F(\theta, \phi; \mathbf{x}) = \sum_{l=0}^{l_{\max}} \sum_{m=-l}^l c_{lm}(\mathbf{x}) Y_l^m(\theta, \phi),
\]
where $(\theta, \phi)$ are the elevation and azimuth angles on the sphere from the initial propagation direction, and $Y_l^m$ are the real SH basis functions.
The peak fiber orientation $\mathbf{d}$ at $\mathbf{x}$ is identified by solving the optimization problem:
\[
\mathbf{d} = \arg\max_{\mathbf{v} \in \mathbb{S}^2} F(\mathbf{v}),
\]
using a Newton--Raphson gradient ascent procedure on the sphere:
\[
\mathbf{v}_{k+1} = \mathbf{v}_k + \delta \mathbf{v},
\]
where $\delta \mathbf{v}$ is computed from the gradient and Hessian of $F(\theta, \phi; \mathbf{x})$ at $\mathbf{v}_k$. 
This optimization typically converges within 5 iterations.

The SD\_Stream algorithm then evaluates the amplitude of the FOD along this orientation:
\[
A = F(\theta_d, \phi_d; \mathbf{x}),
\]
and compares it with a predefined threshold $A_{\text{thresh}}$. 
If $A > A_{\text{thresh}}$, the seed is accepted; otherwise, it is discarded and a new seed point is generated.

For valid seed points, the streamline is updated:
\[
\mathbf{x}_1 = \mathbf{x}_0 + \Delta s \, \mathbf{d}_0,
\]
where $\Delta s$ is the fixed step size, typically much smaller than voxel spacing to prevent overshooting and ensure smooth tracking.

At each subsequent step $n$, propagation proceeds by:
\[
\mathbf{x}_{n+1} = \mathbf{x}_n + \Delta s \, \mathbf{d}_n,
\]
where the next tracking direction $\mathbf{d}_{n}$ is computed at $\mathbf{x}_{n}$ using the same peak-finding procedure.

At every step, termination criteria are checked:
\begin{itemize}
    \item \textbf{Exit image domain} (\texttt{EXIT\_IMAGE}): $\mathbf{x}_{n} \notin \Omega$, where $\Omega$ is the valid image domain.

    \item \textbf{FOD amplitude below threshold} (\texttt{MODEL}): $F(\mathbf{d}_n; \mathbf{x}_n) < A_{\text{thresh}}$.

    \item \textbf{High angular difference between consecutive tracking directions} (\texttt{HIGH\_CURVATURE}): \\ $\cos^{-1}(\mathbf{d}_n \cdot \mathbf{d}_{n-1}) > \theta_{\text{thresh}}$.
    
    \item \textbf{Exit ROI mask} (\texttt{EXIT\_MASK}): $M(\mathbf{x}_{n+1}) = 0$, where $M(\cdot)$ is the ROI mask.
    
    \item \textbf{Reached maximum streamline length d} (\texttt{LENGTH\_EXCEED}): $n \ge N_{\max}$
\end{itemize}

If none of these conditions are met, $\mathbf{x}_{n+1}$ is added as the next point of streamline and propagation continues; otherwise, if any of them are satisfied, propagation terminates.
In bidirectional mode, after forward propagation, tracking resumes from the seed point in the opposite initial direction $-\mathbf{d}_0$, applying the same propagation rules.

The final streamline is constructed by concatenating forward and backward points:
\[
\{\mathbf{x}_{N}, \ldots, \mathbf{x}_0, \ldots, \mathbf{x}_{M}\},
\]
ensuring symmetry around $\mathbf{x}_0$.

Though SD\_Stream is one of the leading streamline propagators in the field, it has rarely been used in end-to- end deep learning frameworks for a few reasons.
First, SD\_Stream is implemented in C++, making it challenging to directly integrate with modern deep learning frameworks, which are predominantly Python-based and rely on automatic differentiation for gradient backpropagation.
Moreover, the original SD\_Stream implementation does not expose a differentiable interface, as it was designed for deterministic streamline propagation rather than optimization within a learning framework.  
The algorithm contains multiple hard-coded control flow operations, such as \texttt{break} and \texttt{return} statements, within its peak-finding and streamline propagation loop.  
These early termination constructs are inherently non-differentiable and interrupt the continuous gradient flow required for end-to-end training in deep learning frameworks.  
Furthermore, SD\_Stream is designed to generate one streamline at a time in a sequential manner, with a variable number of steps depending on local termination conditions, making it difficult to batch-process multiple streamlines efficiently.

\subsection{Differentiable SD\_Stream in PyTorch}
To address the non-differentiable and single-streamline limitations that MRtrix's SD\_Stream have, we present a PyTorch-based implementation of the SD\_Stream propagator that is fully differentiable and capable of processing batched streamlines. 
Our PyTorch implementation closely follows the original C++ logic in MRtrix while redesigning the control flow to eliminate hard-coded operations that disrupt gradient flow, thereby enabling efficient batch-wise propagation.
This PyTorch version of SD\_Stream is specifically designed for seamless integration into deep learning frameworks and supports end-to-end training with backpropagation.

\begin{figure}[h]
    \centering
    \begin{minipage}[t]{0.475\textwidth}
        \small
        \begin{algorithm}[H]
        \caption{Newton–Raphson peak-finding in MRtrix (Single Direction)}
        \label{alg:newton_mrtrix}
        \tcbset{width=7.3cm, colback=light_orange, colframe=black!75, boxrule=0.4pt, arc=2pt, left=2pt, right=2pt, top=1pt, bottom=1pt}
        \begin{tcolorbox}
        \KwIn{FOD volume, SH basis, position $\mathbf{x}$, and initial unit vector $\mathbf{u}$}
        \KwOut{Estimated peak amplitude $A$ at local maximum}

        Interpolate SH coefficients $c$ at $\mathbf{x}$\;
        \end{tcolorbox}
        \vspace*{1\baselineskip}
        \For{$i \leftarrow 1$ \KwTo $50$}{
            \vspace*{1\baselineskip}

            \tcbset{width=6.5cm, colback=light_purple, colframe=black!75, boxrule=0.4pt, arc=2pt, left=2pt, right=2pt, top=1pt, bottom=1pt}
            \begin{tcolorbox}
            Compute azimuth and elevation:\;
            $az \leftarrow \arctan2(u_y, u_x)$ \quad $el \leftarrow \arccos(u_z)$\;
            
            Evaluate $A = \sum_k SH_k(el, az) \cdot c_{k}$\;
            \end{tcolorbox}

            \tcbset{width=6.5cm, colback=light_blue, colframe=black!75, boxrule=0.4pt, arc=2pt, left=2pt, right=2pt, top=1pt, bottom=1pt}
            \begin{tcolorbox}
            Compute gradients $\partial A_x / \partial el_x$, $\partial A_x / \partial az_x$ \;
            Compute gradient norm: \\ 
            $g \leftarrow \sqrt{ (\partial A/\partial el)^2 + (\partial A/\partial az)^2 }$\;

            Normalize gradient: \\
            $d_{el} \leftarrow \partial A/\partial el / g$, \quad
            $d_{az} \leftarrow \partial A/\partial az / g$\; 
            
            Compute $dA/dt$ and $d^2A/dt^2$\;
            \end{tcolorbox}

            \tcbset{width=6cm, colback=light_green, colframe=black!75, boxrule=0.4pt, arc=2pt, left=2pt, right=2pt, top=1pt, bottom=1pt}
            \begin{tcolorbox}
            Compute step: $dt \leftarrow \left| -\frac{dA/dt}{d^2A/dt^2} \right|$\;
            Clamp $dt \leq \texttt{MAX\_DIR\_CHANGE}$\;

            Compute update: \\
            $\delta el = d_{el} dt$, $\delta az = d_{az} dt$\;

            Update $\mathbf{u}$ using spherical increment\;
            Normalize $\mathbf{u}$\;
            \end{tcolorbox}

            \tcbset{width=6cm, colback=yellow, colframe=red, boxrule=2pt, arc=2pt, left=2pt, right=2pt, top=1pt, bottom=1pt}
            \begin{tcolorbox}
            \If{$dt <$\texttt{ANGLE\_TOLERANCE}}{
                \textcolor{red}{\Return} $A$
            }
            \end{tcolorbox}
        }

        Set $\mathbf{u} \leftarrow$ NaN, return NaN\;
        \vspace*{1.3\baselineskip}
        \end{algorithm}
    \end{minipage}
    \hfill
    \begin{minipage}[t]{0.51\textwidth}
        \small
        \begin{algorithm}[H]
        \caption{Newton–Raphson peak-finding in PyTorch (Batched Directions)}
        \label{alg:newton_pytorch}
        \tcbset{width=7.3cm, colback=light_orange, colframe=black!75, boxrule=0.4pt, arc=2pt, left=2pt, right=2pt, top=1pt, bottom=1pt}
        \begin{tcolorbox}
        \KwIn{FOD volume, SH basis, positions $\mathbf{x}$, and directions $\mathbf{d}$}
        \KwOut{Estimated peak amplitude $A$ at local maximum and its directions $\mathbf{d}$}

        Interpolate SH coefficients $c$ at $\mathbf{x}$\;
        \end{tcolorbox}
        Initialize update mask \textcolor{blue}{$m_x$} $\leftarrow 1$\;

        \For{$i \leftarrow 1$ \KwTo $50$}{
            For each $x$ in batch:\;
            \tcbset{width=7.2cm, colback=light_purple, colframe=black!75, boxrule=0.4pt, arc=2pt, left=2pt, right=2pt, top=1pt, bottom=1pt}
            \begin{tcolorbox}
            Compute azimuth and elevation:\;
            $az_x \leftarrow \arctan2(d_{y,x}, d_{x,x})$, \quad $el_x \leftarrow \arccos(d_{z,x})$\;
            
            Evaluate $A_x = \sum_k SH_k(el_x, az_x) \cdot c_{x,k}$\;
            \end{tcolorbox}

            \tcbset{width=7.7cm, colback=light_blue, colframe=black!75, boxrule=0.4pt, arc=2pt, left=2pt, right=2pt, top=1pt, bottom=1pt}
            \begin{tcolorbox}
            Compute gradients $\partial A_x / \partial el_x$, $\partial A_x / \partial az_x$\;

            Compute gradient norm: \\ 
            $g_x = \sqrt{(\partial A/\partial el)^2 + (\partial A/\partial az)^2}$\;

            Normalize gradient: \\
            $\hat{g}_{el,x} = \partial A/\partial el / (g_x + \epsilon)$, \quad
            $\hat{g}_{az,x} = \partial A/\partial az / (g_x + \epsilon)$\;
            
            Compute $dA/dt_x$ and $d^2A/dt^2_x$ using autograd\;
            \end{tcolorbox}

            \tcbset{width=6.7cm, colback=light_green, colframe=black!75, boxrule=0.4pt, arc=2pt, left=2pt, right=2pt, top=1pt, bottom=1pt}
            \begin{tcolorbox}
            Compute step: $dt_x = \left| -\frac{dA/dt_x}{d^2A/dt^2_x} \right|$\;
            Clamp $dt_x \leq \texttt{MAX\_DIR\_CHANGE}$\;

            Compute update: \\
            $\delta el_x = \hat{g}_{el,x} dt_x$ \textcolor{blue}{$m_x$}, \quad
            $\delta az_x = \hat{g}_{az,x} dt_x$ \textcolor{blue}{$m_x$}\;

            Update $\mathbf{d}_x$ using spherical increment\;
            Normalize $\mathbf{d}_x$\;
            \end{tcolorbox}
            \tcbset{width=6.5cm, colback=yellow, colframe=blue, boxrule=2pt, arc=2pt, left=2pt, right=2pt, top=1pt, bottom=1pt}
            \begin{tcolorbox}
            Update mask: \\ \textcolor{blue}{$m_x$} $= 1$ if $dt_x \geq$ \texttt{ANGLE\_TOLERANCE}, else 0\;
            \end{tcolorbox}
        }

        Evaluate $A$ at updated $\mathbf{d}_x$\;
        \textcolor{blue}{\Return} $\mathbf{d}$, $A$\;
        \end{algorithm}
        \vspace*{0.5\baselineskip}
    \end{minipage}
    \caption{Peak-finding using the Newton–Raphson gradient procedure in MRtrix (left) and PyTorch (right). Unlike the MRtrix implementation, which processes a single direction and terminates early when a stopping criterion is met (\textcolor{red}{\textbf{return}} in \textbf{for} loop), the PyTorch version handles multiple directions simultaneously without early termination (\textcolor{blue}{\textbf{return}} out of \textbf{for} loop) through stopping propagation of terminated directions via an update mask \textcolor{blue}{$m_x$}.}
    \label{alg:newton}
\end{figure}

First, we implemented batched trilinear interpolation and spherical harmonic basis evaluation in PyTorch.  
We then extended the Newton–Raphson gradient ascent procedure to operate in batch mode.  
Unlike MRtrix, where peak-finding terminates early when the computed step size falls below a fixed threshold \texttt{ANGLE\_TOLERANCE} (as shown in Algorithm~\ref{alg:newton_mrtrix}), our PyTorch implementation performs a fixed number of 50 iterations, the maximum number of iterations for MRtrix, to ensure differentiability (as shown in Algorithm~\ref{alg:newton_pytorch}).  
Within this iterative procedure, directions that meet the convergence criterion (step size below \texttt{ANGLE\_TOLERANCE}) cease updating, while the remaining directions continue to be refined using the binary update mask until the iteration limit is reached.

\begin{figure}[ht]
    \centering
    \begin{minipage}[t]{0.48\textwidth}
        \small
        \begin{algorithm}[H]
        \caption{MRtrix SD\_Stream Propagation (Single Streamline)}
        \label{alg:prop_mrtrix}
        \KwIn{Seed point $\mathbf{x}_0$, ROI mask, FOD volume}
        \KwOut{Streamline $\mathbf{S}$}
        \vspace*{1\baselineskip}
        Sample initial direction $\mathbf{d}$ randomly from the unit sphere\;
        Compute local FOD maxima $A$ at $\mathbf{x}_0$ and refine $\mathbf{d}$ using Newton–Raphson\;

        \If{$A < A_{\text{thres}}$}{
                \textcolor{red}{\textbf{break}}
            }
        
        Set current position $\mathbf{x} \leftarrow \mathbf{x}_0$\;
        Initialize streamline $\mathbf{S} \leftarrow [\mathbf{x}]$\;
        \vspace*{1\baselineskip}
        \While{streamline length is below max\_length}{

            \tcbset{width=4.5cm, colback=yellow, colframe=red, boxrule=2pt, arc=2pt, left=2pt, right=2pt, top=0.5pt, bottom=0.5pt}
            \begin{tcolorbox}
            \If{$\mathbf{x}$ is out-of-bounds}{
                \textcolor{red}{\textbf{break}}
            }
            \end{tcolorbox}

            \tcbset{width=7.2cm, colback=light_green, colframe=black!75, boxrule=0.4pt, arc=2pt, left=2pt, right=2pt, top=0.5pt, bottom=0.5pt}
            \begin{tcolorbox}
            Update $\mathbf{d_{prev}} \leftarrow \mathbf{d}$\; 
            Update $\mathbf{d}$ using Newton-Raphson\;
            Compute FOD amplitude $A$ at $\mathbf{x}$ in direction $\mathbf{d}$\;
            \end{tcolorbox}

            \tcbset{width=4.5cm, colback=yellow, colframe=red, boxrule=2pt, arc=2pt, left=2pt, right=2pt, top=0.5pt, bottom=0.5pt}
            \begin{tcolorbox}
            \If{$A < A_{\text{thres}}$}{
                \textcolor{red}{\textbf{break}}
            }
            \end{tcolorbox}
            
            \tcbset{width=4.5cm, colback=yellow, colframe=red, boxrule=2pt, arc=2pt, left=2pt, right=2pt, top=0.5pt, bottom=0.5pt}
            \begin{tcolorbox}
            \If{$\angle(\mathbf{d}, \mathbf{d_{prev}}) > \theta_{\text{thres}}$}{
                \textcolor{red}{\textbf{break}}
            }
            \end{tcolorbox}
            
            \tcbset{width=3.6cm, colback=light_orange, colframe=black!75, boxrule=0.4pt, arc=2pt, left=2pt, right=2pt, top=0.5pt, bottom=0.5pt}
            \begin{tcolorbox}
            Update $\mathbf{x} \leftarrow \mathbf{x} + s \cdot \mathbf{d}$\;
            \end{tcolorbox}

            \tcbset{width=4.5cm, colback=yellow, colframe=red, boxrule=2pt, arc=2pt, left=2pt, right=2pt, top=0.5pt, bottom=0.5pt}
            \begin{tcolorbox}
            \If{$\textbf{x}$ is in invalid ROI}{
                \textcolor{red}{\textbf{break}}
            }
            \end{tcolorbox}
            Append $\mathbf{x}$ to $\mathbf{S}$\;
        }
        \vspace*{1\baselineskip}
        \Return{$\mathbf{S}$}
        \vspace*{2\baselineskip}
        \end{algorithm}
    \end{minipage}
    \hfill
    \begin{minipage}[t]{0.48\textwidth}
        \small 
        \begin{algorithm}[H]
        \caption{PyTorch SD\_Stream Propagation (Batched Streamlines)}
        \label{alg:prop_torch}
        \KwIn{Seed points $\mathbf{x}_0$, initial directions $\mathbf{d}_0$, ROI mask, FOD volume}
        \KwOut{Streamlines $\mathbf{S}$ and valid lengths $\mathbf{L}$}
        Compute local FOD maxima $A_0$ at $\mathbf{x}_0$ and refine $\mathbf{d}_0$ using Newton–Raphson\;
        Initialize streamline tensor $\mathbf{S} \leftarrow 0$\;
        Set $\mathbf{S}[:, 0, :] \leftarrow \mathbf{x}_0$\;
        Initialize active mask \textcolor{blue}{$M$} $\leftarrow \mathbf{1}$\;
        Initialize valid length vector $\mathbf{L} \leftarrow -1$\;
        \For{$t \leftarrow 0$ \KwTo max\_length $- 1$}{
            \tcbset{width=6cm, colback=yellow, colframe=blue, boxrule=2pt, arc=2pt, left=2pt, right=2pt, top=0.5pt, bottom=0.5pt}
            \begin{tcolorbox}
            \If{$\mathbf{x_t}$ is out-of-bound}{
                \textcolor{blue}{\textbf{Deactivate streamline}} (\textcolor{blue}{$M$} $\leftarrow 0$)
            }
            \end{tcolorbox}

            \tcbset{width=6cm, colback=yellow, colframe=blue, boxrule=2pt, arc=2pt, left=2pt, right=2pt, top=0.5pt, bottom=0.5pt}
            \begin{tcolorbox}
            \If{$A_t < A_{\text{thres}}$}{
                \textcolor{blue}{\textbf{Deactivate streamline}} (\textcolor{blue}{$M$} $\leftarrow 0$)
            }
            \end{tcolorbox}
            
            \tcbset{width=6.1cm, colback=yellow, colframe=blue, boxrule=2pt, arc=2pt, left=2pt, right=2pt, top=0.5pt, bottom=0.5pt}
            \begin{tcolorbox}
            \If{$t > 0$ and $\angle(\mathbf{d}_{t-1}, \mathbf{d}_t) > \theta_{\text{thres}}$}{
                \textcolor{blue}{\textbf{Deactivate streamline}} (\textcolor{blue}{$M$} $\leftarrow 0$)
            }
            \end{tcolorbox}
            
            \tcbset{width=4.5cm, colback=light_orange, colframe=black!75, boxrule=0.4pt, arc=2pt, left=2pt, right=2pt, top=0.5pt, bottom=0.5pt}
            \begin{tcolorbox}
            Propagate to next position: \\ $\mathbf{x}_{t+1} = \mathbf{x}_t + \text{step\_size} \cdot \mathbf{d}_t$\;
            \end{tcolorbox}

            \tcbset{width=5cm, colback=light_green, colframe=black!75, boxrule=0.4pt, arc=2pt, left=2pt, right=2pt, top=0.5pt, bottom=0.5pt}
            \begin{tcolorbox}
            At $\mathbf{x}_{t+1}$, update $\mathbf{d}_{t+1}$ and $A_{t+1}$ \\ using Newton–Raphson\;
            \end{tcolorbox}

            \tcbset{width=6cm, colback=yellow, colframe=blue, boxrule=2pt, arc=2pt, left=2pt, right=2pt, top=0.5pt, bottom=0.5pt}
            \begin{tcolorbox}
            \If{$\mathbf{x_t}$ is in invalid ROI}{
                \textcolor{blue}{\textbf{Deactivate streamline}} (\textcolor{blue}{$M$} $\leftarrow 0$)
            }
            \end{tcolorbox}
            Set $\mathbf{S}[\textcolor{blue}{M}, t+1, :] \leftarrow \mathbf{x}_{t+1}[\textcolor{blue}{M}]$\;
            \tcbset{width=7cm, colback=light_green, colframe=black!75, boxrule=0.4pt, arc=2pt, left=2pt, right=2pt, top=0.5pt, bottom=0.5pt}
            \begin{tcolorbox}
            Update $(\mathbf{x}_t, \mathbf{d}_t, A_t) \leftarrow (\mathbf{x}_{t+1}, \mathbf{d}_{t+1}, A_{t+1})$ for active streamlines\;
            \end{tcolorbox}
            Set $\mathbf{L}[i] \leftarrow t+1$ for each newly deactivated streamline $i$\;
        }
        \Return{$\mathbf{S}$ and $\mathbf{L}$}
        \end{algorithm}
    \end{minipage}
    \caption{SD\_Stream algorithm in MRtrix (left) and PyTorch (right). Unlike the MRtrix implementation, which processes a single streamline and terminates early (using \textcolor{red}{\textbf{break}})  when a single stopping criterion is met, the PyTorch version handles multiple streamlines simultaneously without early termination by utilizing a masking procedure (\textcolor{blue}{$M$}) to track termination status.}
    \label{fig:prop}
\end{figure}

Next, we implemented a batched version of the SD\_Stream algorithm (Algorithm~\ref{alg:prop_torch}) using the PyTorch-based batched Newton-Raphson gradient ascent procedure.  
As in MRtrix (Algorithm~\ref{alg:prop_mrtrix}), we evaluated termination criteria at each step, including \texttt{EXIT\_IMAGE}, \texttt{MODEL}, \texttt{HIGH\_CURVATURE}, \texttt{EXIT\_MASK}, and \texttt{LENGTH\_EXCEED}.  
However, unlike MRtrix's streamline propagation, our implementation does not immediately terminate streamline propagation when these conditions are met.  
Instead, propagation continues up to a predefined maximum number of points.
Streamlines that meet termination criteria are padded with zeros from the point of termination onward, guided by a binary active mask.  
As a result, the output streamlines from our PyTorch implementation represent zero-padded versions of the valid streamlines.

We implemented our streamline propagator in a CPU-based setting, as the accumulation of gradient values—necessary for maintaining differentiability—leads to substantial memory overhead on the GPU, making CPU execution more efficient in this case.
However, the entire process can be transferred to the GPU, allowing all operations to run within a CUDA environment.

With this batched and non-early-stopping design, our streamline propagator is fully differentiable and capable of processing multiple streamlines simultaneously while maintaining compatibility with gradient-based optimization.
This design ensures that all streamlines in a batch can be processed uniformly, allowing the entire streamline propagation procedure to be incorporated as a differentiable module within deep learning workflows.

\subsection{Datasets and Data Preprocessing}
To validate our PyTorch implemented SD\_Stream against MRtrix's original SD\_Stream, we used a dMRI scan of an adult participant from the Human Connectome Project (HCP) \cite{van2012human}.
The scanner used to acquire the dMRI data was a 3T Siemens Skyra (Erlangen, Germany), with a multishell single-shot echo planar imaging acquisition at b = 1000, 2000, and 3000 \(\mathrm{s/mm^2}\) and 90 directions per shell (\(\mathrm{TE/TR = 89.5/5520ms}\)) with 6 b=0 \(\mathrm{s/mm^2}\) images. 
The dMRI data was susceptibility, motion and eddy current corrected.
We then estimated the FODs using the CSD algorithm \cite{tournier2007robust} and generated 10,000 streamlines using the \texttt{SD\_Stream} algorithm (\texttt{tckgen} command) on the estimated FODs, both implemented in MRtrix \cite{tournier2019mrtrix3}.
Seed points were uniformly sampled within the white matter, and tracking was constrained to remain within a brain mask.
A fixed step size of 1 mm was used, and streamlines were constrained to lengths between 50 mm and 100 mm.
Tracking was performed in unidirectional mode and the coordinates of all seed points were saved.

\subsection{Experiments}
To reproduce streamlines generated by MRtrix’s SD\_Stream algorithm using our PyTorch implementation, we extracted the initial tracking directions and paired them with the corresponding seed points.
This step is essential because, in MRtrix, initial directions are randomly sampled from the sphere and different initializations can lead to substantially different streamline trajectories.
We then used our PyTorch implementation of SD\_Stream, giving the same FODs used as input for the original MRtrix version of SD\_Stream, paired initial directions and seed points to generate streamlines.
To evaluate the agreement between the streamlines generated by the two methods, we computed the Hausdorff distance between the corresponding streamlines, measuring the greatest distance from a point on one streamline to the closest point on the other.

We also measured the runtime and memory usage when generating 2,000 streamlines for both methods, as well as the time and additional memory required for gradient backpropagation through a single streamline.
All experiments were conducted using MRtrix3 version 3.4.0 and PyTorch version 2.6.0 on a 32 core CPU system with 192 GB of RAM.

\section{Results}

\begin{table}[h]
\begin{center}
\begin{tabular}{|c|cc|cc|}
\hline
\multirow{2}{*}{\textbf{Methods}} & \multicolumn{2}{c|}{\textbf{Generating 2,000 streamlines}}           & \multicolumn{2}{c|}{\textbf{Propagating gradients in one streamline}} \\ \cline{2-5} 
                                  & \multicolumn{1}{c|}{\textbf{Time (seconds)}} & \textbf{Memory (MB)} & \multicolumn{1}{c|}{\textbf{Time (seconds)}}  & \textbf{Memory (MB)}  \\ \hline
\textbf{MRtrix}                   & \multicolumn{1}{c|}{1.01}                     & 372                  & \multicolumn{1}{c|}{Not available}             & Not available         \\ \hline
\textbf{PyTorch}                  & \multicolumn{1}{c|}{930}                      & 89,600               & \multicolumn{1}{c|}{111.5}                     & 0                     \\ \hline
\end{tabular}
\end{center}
\caption{Comparison of runtime and memory usage for generating 2,000 streamlines using MRtrix and PyTorch, along with the time and additional memory required for gradient backpropagation through a single streamline. 
While PyTorch requires significantly more time and memory to generate the same number of streamlines compared to MRtrix, it supports gradient backpropagation, a feature that is not available in MRtrix.
}
\label{table:time}
\end{table}

Table~\ref{table:time} shows the runtime and memory usage for generating 2,000 streamlines using the MRtrix and PyTorch implementations, as well as the time and memory overhead associated with gradient backpropagation through a single streamline.
It is evident that the MRtrix is noticeably more efficient for generating the same number of streamlines, being approximately 920 times faster and using 240 times less memory than the PyTorch implementation.
However, unlike MRtrix, the PyTorch version supports gradient backpropagation to the input FODs, an important capability for integrating streamline generation into end-to-end deep learning pipelines.

\begin{figure}[ht]
    \centering
    \includegraphics[width=0.99\linewidth]{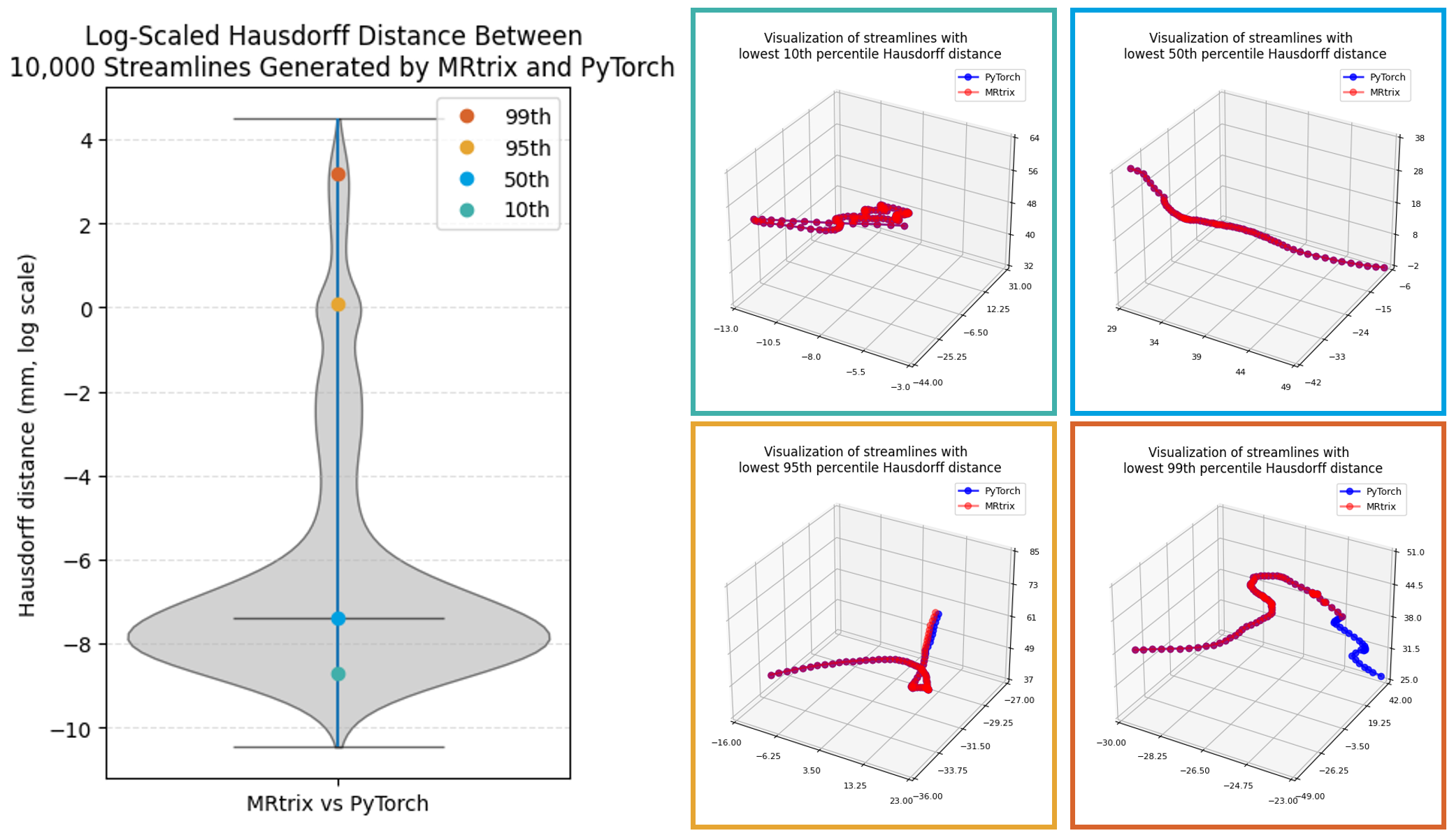}
    \caption{Violin plot of the log-scaled Hausdorff distances between 10,000 streamlines generated by MRtrix and PyTorch. 
    Colored markers indicate the 10th, 50th, 95th, and 99th percentiles.
    The corresponding streamlines at each percentile are visualized on the right, comparing spatial alignment between the PyTorch (blue) and MRtrix (red) methods.}
    \label{fig:haus}
\end{figure}

Figure~\ref{fig:haus} shows the distribution of log-scaled Hausdorff distances between 10,000 streamlines generated by MRtrix and our PyTorch implementation, along with visualizations of representative streamline pairs corresponding to the 10th, 50th, 95th, and 99th percentiles.
The results indicate that our PyTorch implemented SD\_Stream accurately reproduces streamlines that well-aligned with those from MRtrix, with more than 9,300 streamlines having a Hausdorff distance below 1 mm.

Streamline pairs at the 10th and 50th percentiles demonstrate near-perfect alignment, with Hausdorff distances close to 0.0 mm and 0.001 mm, respectively.
However, streamline pairs in the 95th and 99th percentiles exhibit visible misalignments near the posterior end, with Hausdorff distances reaching approximately 1.073 mm and 24.25 mm, respectively.
In the 95th percentile case, two streamlines branch out from a specific point. 
This divergence originates from accumulated floating-point differences between C++ (used in MRtrix) and Python (used in PyTorch). 
These discrepancies result in slightly different FOD values at the corresponding locations, ultimately causing a mismatch in streamline termination steps, with PyTorch terminating one step earlier compared to MRtrix propagation.

In the 99th percentile case, while the streamlines are largely aligned, the PyTorch streamline continues propagating after the MRtrix streamline has terminated. 
This is due to the early stopping rule in the peak-finding procedure of the Newton–Raphson ascent, where an update step exceeding the \texttt{ANGLE\_TOLERANCE} threshold is treated as non-convergent, prompting the termination of the tracking.
This update step is computed using the gradient and Hessian of the FOD amplitude with respect to direction, and may differ inherently between the C++ and Python implementations due to fundamental differences in numerical precision and floating-point operations.
This type of discrepancy was observed in only 47 out of 10,000 streamlines. 
Considering that in typical deep learning applications, the PyTorch-generated streamlines are cropped to match the length of the corresponding MRtrix streamlines before comparison, such deviations are unlikely to have a significant impact on model performance.

\section{Discussion}
In this work, we present a fully differentiable streamline propagator that allows uninterrupted gradient flow throughout the entire pipeline.
Our streamline propagator is a PyTorch implementation of the SD\_Stream algorithm from MRtrix, originally written in C++, with modifications to remove components that block gradient flow, making it fully differentiable.
Although the current version runs on the CPU to manage gradient-related memory costs, it remains fully portable to GPU for CUDA environment if needed.
Our PyTorch-based implementation demonstrates strong numerical consistency with the streamlines generated from MRtrix's SD\_Stream, achieving over 93\% of streamlines with a Hausdorff distance below 1 mm out of 10,000 streamlines. 
Although our method is computationally intensive—being approximately 920 times slower and consuming 240 times more memory than MRtrix, it is fully differentiable, supports batch processing of multiple streamlines, and is seamlessly compatible with modern deep learning frameworks. 
This makes it particularly suitable for integration into end-to-end training pipelines. 
Our differentiable streamline propagator provides a foundation for incorporating macrostructural reasoning into deep learning-based tractography and related applications.

\acknowledgments 
This work was conducted in part using the resources of the Advanced Computing Center for Research and Education (ACCRE) at Vanderbilt University, Nashville, TN.
The Vanderbilt Institute for Clinical and Translational Research (VICTR) is funded by the National Center for Advancing Translational Sciences (NCATS) Clinical Translational Science Award (CTSA) Program, Award Number 5UL1TR002243-03.
This work was supported by the National Institute of Health (NIH) under award number 5U01DA055347-03 and the Alzheimer’s Disease Sequencing Project Phenotype Harmonization Consortium (ADSP-PHC) that is funded by the National Institute of Aging (NIA) (U24 AG074855, U01 AG068057 and R01 AG059716).
The content is solely the responsibility of the authors and does not necessarily represent the official views of the NIH or the NIA.

We used generative artificial intelligence (AI) to create code segments based on task descriptions, as well as to debug, edit, and autocomplete code. Additionally, generative AI technologies have been employed to assist in structuring sentences and performing grammatical checks. The conceptualization, ideation, and all prompts provided to the AI originated entirely from the authors’ creative and intellectual efforts. We take accountability for the review of all content generated by AI in this work.

\bibliography{report} 
\bibliographystyle{spiebib} 

\end{document}